\documentstyle[preprint,aps]{revtex}
\newread\epsffilein    % file to \read
\newif\ifepsffileok    % continue looking for the bounding box?
\newif\ifepsfbbfound   % success?
\newif\ifepsfverbose   % report what you're making?
\newdimen\epsfxsize    % horizontal size after scaling
\newdimen\epsfysize    % vertical size after scaling
\newdimen\epsftsize    % horizontal size before scaling
\newdimen\epsfrsize    % vertical size before scaling
\newdimen\epsftmp      % register for arithmetic manipulation
\newdimen\pspoints     % conversion factor
\def\epsfbox#1{\global\def\epsfllx{72}\global\def\epsflly{72}%
   \global\def\epsfurx{540}\global\def\epsfury{720}%
   \def\lbracket{[}\def\testit{#1}\ifx\testit\lbracket
   \let\next=\epsfgetlitbb\else\let\next=\epsfnormal\fi\next{#1}}%
\def\epsfgetlitbb#1#2 #3 #4 #5]#6{\epsfgrab #2 #3 #4 #5 .\\%
   \epsfsetgraph{#6}}%
\def\epsfnormal#1{\epsfgetbb{#1}\epsfsetgraph{#1}}%
\def\epsfgetbb#1{%
\openin\epsffilein=#1
\ifeof\epsffilein\errmessage{I couldn't open #1, will ignore it}\else
   {\epsffileoktrue \chardef\other=12
    \def\do##1{\catcode`##1=\other}\dospecials \catcode`\ =10
    \loop
       \read\epsffilein to \epsffileline
       \ifeof\epsffilein\epsffileokfalse\else
          \expandafter\epsfaux\epsffileline:. \\%
       \fi
   \ifepsffileok\repeat
   \ifepsfbbfound\else
    \ifepsfverbose\message{No bounding box comment in #1; using defaults}\fi\fi
   }\closein\epsffilein\fi}%
\def\epsfclipstring{}% do we clip or not?  If so,
\def\epsfsetgraph#1{%
   \epsfrsize=\epsfury\pspoints
   \advance\epsfrsize by-\epsflly\pspoints
   \epsftsize=\epsfurx\pspoints
   \advance\epsftsize by-\epsfllx\pspoints
   \epsfxsize\epsfsize\epsftsize\epsfrsize
   \ifnum\epsfxsize=0 \ifnum\epsfysize=0
      \epsfxsize=\epsftsize \epsfysize=\epsfrsize
      \epsfrsize=0pt
     \else\epsftmp=\epsftsize \divide\epsftmp\epsfrsize
       \epsfxsize=\epsfysize \multiply\epsfxsize\epsftmp
       \multiply\epsftmp\epsfrsize \advance\epsftsize-\epsftmp
       \epsftmp=\epsfysize
       \loop \advance\epsftsize\epsftsize \divide\epsftmp 2
       \ifnum\epsftmp>0
          \ifnum\epsftsize<\epsfrsize\else
             \advance\epsftsize-\epsfrsize \advance\epsfxsize\epsftmp \fi
       \repeat
       \epsfrsize=0pt
     \fi
   \else \ifnum\epsfysize=0
     \epsftmp=\epsfrsize \divide\epsftmp\epsftsize
     \epsfysize=\epsfxsize \multiply\epsfysize\epsftmp   
     \multiply\epsftmp\epsftsize \advance\epsfrsize-\epsftmp
     \epsftmp=\epsfxsize
     \loop \advance\epsfrsize\epsfrsize \divide\epsftmp 2
     \ifnum\epsftmp>0
        \ifnum\epsfrsize<\epsftsize\else
           \advance\epsfrsize-\epsftsize \advance\epsfysize\epsftmp \fi
     \repeat
     \epsfrsize=0pt
    \else
     \epsfrsize=\epsfysize
    \fi
   \fi
   \ifepsfverbose\message{#1: width=\the\epsfxsize, height=\the\epsfysize}\fi
   \epsftmp=10\epsfxsize \divide\epsftmp\pspoints
   \vbox to\epsfysize{\vfil\hbox to\epsfxsize{%
      \ifnum\epsfrsize=0\relax
        \includegraphics{#1}%
      \else
        \epsfrsize=10\epsfysize \divide\epsfrsize\pspoints
        \includegraphics{#1}%
      \fi
      \hfil}}%
\global\epsfxsize=0pt\global\epsfysize=0pt}%
{\catcode`\%=12 \global\let\epsfpercent=%\global\def\epsfbblit{%BoundingBox}}%
\long\def\epsfaux#1#2:#3\\{\ifx#1\epsfpercent
   \def\testit{#2}\ifx\testit\epsfbblit
      \epsfgrab #3 . . . \\%
      \epsffileokfalse
      \global\epsfbbfoundtrue
   \fi\else\ifx#1\par\else\epsffileokfalse\fi\fi}%
\def\epsfempty{}%
\def\epsfgrab #1 #2 #3 #4 #5\\{%
\global\def\epsfllx{#1}\ifx\epsfllx\epsfempty
      \epsfgrab #2 #3 #4 #5 .\\\else
   \global\def\epsflly{#2}%
   \global\def\epsfurx{#3}\global\def\epsfury{#4}\fi}%
\def\epsfsize#1#2{\epsfxsize}
\let\epsffile=\epsfbox

\begin{document}
\draft
\title{An exchange-correlation energy for a two- \\
dimensional electron gas in a magnetic field}
\author{Rodney Price and S.\ Das Sarma}
\address{Dept.\ of Physics, University of Maryland, College Park MD
20742}
\maketitle
\begin{abstract}
We present the results of a variational Monte Carlo calculation of the
exchange-correlation energy for a spin-polarized two-dimensional
electron gas in a perpendicular magnetic field.  These energies are a
necessary input to the recently developed current-density functional
theory.  Landau-level mixing is included in a variational manner, which
gives the energy at finite density at finite field, in contrast to
previous approaches.  Results are presented for the
exchange-correlation energy and excited-state gap at $\nu =$ 1/7, 1/5,
1/3, 1, and 2.  We parameterize the results as a function of $r_s$ and
$\nu$ in a form convenient for current-density functional calculations.
\end{abstract}

\pacs{ }

\narrowtext
\newpage

\section{Introduction}

Two-dimensional electron systems have been a subject of great interest
for some time now, as fascinating phenomena such as the fractional
quantum Hall effect and the Wigner solid have been studied intensively.
 More recently, technological progress has allowed experimenters to
fabricate extremely small structures, quantum dots, wires, and so on,
that function as external potentials imposed on the 2-D electron gas.
An extension of density functional theory, current density functional
theory (CDFT) \cite{vignale87,vignale88}, has been used to study these
phenomena \cite{vignale93,quantumdot}.  The use of CDFT in this context
is attractive, but the calculations to date have been hampered by the
lack of an accurate exchange-correlation energy.  The ones in use have
relied on an interpolation between results at zero magnetic field as a
function of density \cite{tanatar89,ceperley78}, and at infinite
magnetic field and infinite density \cite{levesque84}; i.e.\ the lowest
Landau level approximation.  The CDFT uses derivatives of the
exchange-correlation energy with respect to density and magnetic field,
so these interpolations will reflect only zero- and infinite- magnetic
field and density properties of the system.  In addition, since the
lowest Landau level approximation requires that the density $n$ and
magnetic field $B$ go to infinity with the ratio $\sqrt{n}/B$ held
constant, a degree of freedom in the theory is lost.

In this paper, we report the results of variational quantum Monte Carlo
calculations of ground-state energies at various integer and fractional
magnetic filling factors $\nu$.  Our results include an estimate of
Landau-level mixing, which allows us to calculate these quantities at
finite density and magnetic field.  In addition, we find the
quasielectron-quasihole gaps at fractional filling factors, including
the effect of Landau level mixing here as well.  Using both these
quantities yields a picture of the true exchange-correlation potential
near $\nu$ at varying density, including the FQHE-induced cusp.

The CDFT generalizes the usual density functional theory by including
the coupling of orbital currents to a magnetic field.  The functionals
in the theory depend not only on the particle density $n(\vec{r})$ but
on the paramagnetic current density $\vec{j}_p(\vec{r})$ as well.  (The
physical current density $\vec{j}(\vec{r})$ is obtained using the
continuity equation.)  Solving the CDFT Kohn-Sham equations
\cite{vignale88} for the single-particle orbitals $\psi(\vec{r})$ then
lets us define $n(\vec{r})$ and $\vec{j}_p(\vec{r})$ self-consistently,
\begin{eqnarray}
n(\vec{r}) & = & \sum_i | \psi_i(\vec{r}) |^2  \\
\vec{j}_p(\vec{r}) & = & \frac{-i\hbar}{2m} \sum_i \left[
\psi^*_i(\vec{r}) \nabla \psi_i(\vec{r}) - \psi_i(\vec{r}) \nabla
\psi^*_i(\vec{r}) \right].
\end{eqnarray}

The Kohn-Sham equations involve two exchange-correlation potentials: a
scalar potential $V_{xc}(\vec{r})$ and a vector potential
$\vec{A}_{xc}(\vec{r})$.  These potentials, in turn, are functionals of
the exchange-correlation energy $E_{xc}$ as follows,
\begin{eqnarray}
V_{xc}(\vec{r}) & = & \frac{\delta E_{xc}[ n(\vec{r}), \vec{v}(\vec{r})
]}{\delta n(\vec{r})} \\
\vec{A}_{xc}(\vec{r}) & = & - \frac{c}{n(\vec{r})} \nabla \times \left(
\frac{\delta E_{xc}[ n(\vec{r}), \vec{v}(\vec{r}) ]}{\delta
\vec{v}(\vec{r})} \right),
\end{eqnarray}
where we have defined the vorticity
\begin{equation}
\vec{v}(\vec{r}) = \nabla \times \left(
\frac{\vec{j}_p(\vec{r})}{n(\vec{r})} \right).
\end{equation}
The fact that $\vec{j}_p(\vec{r})$ enters in only through the vorticity
$\vec{v}(\vec{r})$ ensures that the potentials are gauge-invariant.  In
addition, a local current density approximation (LCDA) can be defined
in terms of the density $n(\vec{r})$ and the vorticity
$\vec{v}(\vec{r})$:
\begin{equation}
E_{xc}[n(\vec{r}), \vec{v}(\vec{r})] = \int d^2r\, n(\vec{r})
\varepsilon_{xc}( n(\vec{r}), \vec{v}(\vec{r}) ).
\end{equation}

As with the LDA, this local energy $\varepsilon_{xc}$ is the
exchange-correlation energy of a uniform state, in this case the FQHE
liquid state.  Because the magnetic field in this state is uniform,
translational invariance requires that the physical current density
$\vec{j}(\vec{r})$ must be uniform and therefore zero.  Then
\begin{equation}
\vec{j}(\vec{r}) = \vec{j}_p(\vec{r}) - \frac{e n_0}{mc}
\vec{A}(\vec{r}) = 0,
\end{equation}
where $\vec{A}(\vec{r})$ is the physical vector potential and $n_0$ is
the density.  Taking the curl of both sides yields
\begin{equation}
\nabla \times \left( \frac{\vec{j}_p(\vec{r})}{n_0} \right) =
\frac{e}{mc} \nabla \times \vec{A}(\vec{r}) =
\frac{e}{mc} \vec{B}.
\label{uniform_vorticity}
\end{equation}
The left-hand side of (\ref{uniform_vorticity}) is just the vorticity,
so using the definition of the magnetic filling factor $\nu = 2 \pi n_0
\hbar c/eB$, we have
\begin{equation}
\vec{v}(\vec{r}) = \frac{2 \pi \hbar n_0}{m \nu}.
\end{equation}

The vorticity in the uniform liquid state is then a function of the
density $n_0$ and filling factor $\nu$.  Then in order to use the LCDA,
we need only find the exchange-correlation energy of the uniform system
\begin{equation}
\bar{\varepsilon}_{xc}(r_s, \nu) = \varepsilon_{xc}( n(\vec{r}),
\vec{v}(\vec{r}) )
\end{equation}
as a function of density (or equivalently ion-disk radius $r_s = (\pi
n_0)^{-1/2}$) and filling factor $\nu$.

\section{Variational Monte Carlo method}

We calculate the exchange-correlation energy $\bar{\varepsilon}_{xc}$
using a variational quantum Monte Carlo method.  The method uses the
Metropolis algorithm to evaluate the energy of a given wavefunction at
various values of a variational parameter $\alpha$.  Since we are
interested in including Landau-level mixing in the wavefunction only at
certain fixed values of $\nu$, where the liquid state is
incompressible, we would like to use wavefunctions with suitable
correlations built in already.  The parameter $\alpha$ will be used to
vary the amount of Landau-level mixing in the wavefunction.  We can do
this if we use lowest Landau level (LLL) wavefunctions for the
incompressible states which are well-known: i.e. the Laughlin states
for fractional $\nu$ or Slater determinants for integer $\nu$.  We
obtain our variational wavefunction $\psi_\alpha$ by multiplying the
LLL wavefunction by a Jastrow factor $J_\alpha$ which lifts the
wavefunction partially out of the LLL:
\begin{equation}
J_\alpha = \exp \left[ -\alpha \sum_{i<j} u(R_{ij}) \right],
\end{equation}
where $R_{ij}$ is the distance between the electrons $i$ and $j$, and
$u(r)$ is the so-called pseudopotential.

Now, since $\psi_\alpha = \psi_0 J_\alpha$, where $\psi_0$ is the LLL
wavefunction, varying $\alpha$ varies the amount of Landau-level
mixing. (The Jastrow factor forces $\psi_\alpha$ out of the LLL since
it is non-analytic.) When $\alpha = 0$ we recover the LLL wavefunction.

The pseudopotential $u(r)$ is chosen to reproduce the zero-point motion
of the plasmon modes in two dimensions \cite{ceperley78}; then
\begin{equation}
u(r) \sim \frac{1}{\sqrt{r}}
\label{oldpseudo}
\end{equation}
at large $r$.  This pseudopotential has a singularity at the origin.
In order to minimize the kinetic energy of the wavefunction we should
require $u(r) \rightarrow 1/\sqrt{F}$ and $\frac{du}{dr} \rightarrow
-\frac{1}{3} \left( \frac{1}{F} \right)^{3/2}$ as $r \rightarrow 0$,
where $F$ is another variational parameter.  A pseudopotential
satisfying these constraints is given by
\begin{equation}
u(r) = \frac{1}{\sqrt{r}} \left[ 1 - \exp \left(-\sqrt{\frac{r}{F}} -
\frac{r}{2F} \right) \right].
\label{pseudopotential}
\end{equation}
In practice, we have found that choosing $F \approx 0.5$ such that
$\frac{du}{dr} \rightarrow -\frac{1}{3} \left( \frac{1}{F}
\right)^{3/2} = -1$ minimizes the energy at arbitrary $\alpha$, and
thus $r_s$, quite well.  In all the calculations that follow, we have
used $F = 0.5$, and varied only the parameter $\alpha$.  A comparison
of pseudopotentials (\ref{oldpseudo}) and (\ref{pseudopotential}) is
shown in Figure \ref{pseudopotential_graph}.

Whether our variational wavefunction is the ground state or an excited
state, we wish to find its total energy
\begin{eqnarray}
E_\alpha & = & \left< \alpha | H | \alpha \right> \\
         & = & \frac{1}{r_s^2} \left< \alpha | K | \alpha \right> +
\frac{2}{r_s} \left< \alpha | V | \alpha \right>, \nonumber
\end{eqnarray}
where $H$ is the Hamiltonian, $K$ is the kinetic energy operator, $V$
is the potential energy operator and $\left.| \alpha\right>$ is the
many-particle state.  The ion-disk radius $r_s$ factors completely out
of both kinetic and potential parts.  These two energies, $\left<
\alpha | K | \alpha \right>$ and $\left< \alpha | V | \alpha \right>$,
are computed for a range of $\alpha$. The energy $E(r_s)$ is found at
any given $r_s$ by minimizing $E_\alpha$ with respect to $\alpha$.
This allows us to find $E(r_s)$ in a corresponding range of $r_s$
without recomputing $E_\alpha$ for each new $r_s$.

The potential energy $\left< \alpha | V | \alpha \right>$ is easy to
calculate by Monte Carlo.  We write
\begin{equation}
\left< \alpha | V | \alpha \right> = \int \! d\Omega |\psi_\alpha|^2
V(\Omega)
\end{equation}
where $\Omega \equiv \{\Omega_1, \cdots , \Omega_N\}$ and $\Omega_i$ is
the position of the $i$th electron on the sphere.  The integral is
taken over all $\Omega_i$.  We sample from the probability density
$|\psi_\alpha|^2$ and sum the potential energy
\begin{equation}
V(\Omega) = \sum_{i<j}^N \frac{1}{R_{ij}}.
\label{coulomb}
\end{equation}

The kinetic energy sum is also straightforward, but more
computationally demanding.  The integral becomes
\begin{equation}
\left< \alpha | K | \alpha \right> = \int \! d\Omega |\psi_\alpha|^2
\frac{K\psi_\alpha}{\psi_\alpha}.
\end{equation}
As $\psi_\alpha$ will sometimes involve a projection of an already
complex wavefunction to the LLL, we will have to take pains to make
this computation efficient.  If the projection is necessary, the cost
of computing $(K\psi_\alpha)/\psi_\alpha$ is of order $N^3$, the most
computationally expensive part of the code.

\subsection{Single-particle states and operators}

Our calculations are done on the sphere \cite{fano86} rather than in
the plane, as there are no boundaries and the calculations are more
efficient.  Since the calculations involve Landau-level mixing, which
makes the particle density finite, we choose the radius of the sphere
$R = \sqrt{N}/2$, where $N$ is the number of particles, to keep the
area per particle constant.  The magnetic monopole at the center of the
sphere has magnetic charge $S$.  We use units of length $a = (\pi
n_0)^{-1/2}$ and units of energy $e^2/\epsilon a_B$, where $a_B =
\hbar^2 \epsilon / m^* e^2$ and $r_s = a/a_B$.  The position of an
electron on the sphere is described in spinor coordinates $\chi =
(u,v)$, where
%\begin{eqnarray}
%u & = & e^{-i \phi/2} \cos \frac{\theta}{2} \\
%v & = & e^{i \phi/2} \sin \frac{\theta}{2}
%\end{eqnarray}
\begin{equation}
\left( \begin{array}{c} u \\ v \end{array} \right) =
\left( \begin{array}{c} e^{-i \phi/2} \cos \frac{\theta}{2} \\
                        e^{i \phi/2} \sin \frac{\theta}{2} \end{array}
\right)
\end{equation}
and $(\theta, \phi)$ is the electron position in spherical coordinates.

The kinetic energy of a single particle on the sphere can be described
in terms of its angular momentum \cite{haldane83,fano86}
\begin{equation}
K = \frac{L^2 - S^2}{R^2}.
\label{kinetic_single}
\end{equation}
If we choose the gauge
\begin{equation}
\vec{A} = -\frac{S}{R} \cot \theta \hat{\phi},
\end{equation}
we find that the single-particle states are the monopole harmonics
\cite{wu76}
\begin{eqnarray}
Y_{S,l,m} & = & M_{S,l,m} u^{S-m} v^{S+m} P_{l-s}^{S+m,S-m}(u \bar{u} -
v \bar{v}), \\
M_{S,l,m} & = & \left[ \frac{2l+1}{4\pi}
\frac{(l-S)!(l+S)!}{(l-m)!(l+m)!} \right] ^{1/2}. \nonumber
\end{eqnarray}
Here $l \geq S$ is the Landau-level index, and $P_n^{\alpha, \beta}(x)$
is a Jacobi polynomial.  (A bar over a variable denotes conjugation.)
If we set $l=S$ we recover the LLL states given in \cite{fano86}.

The angular momentum operators given in \cite{fano86} are valid only in
the LLL.  Operators valid for all Landau levels must include the
conjugate variables $\bar{u}$ and $\bar{v}$.  The angular momentum
operators then become
\begin{eqnarray}
L_+ & = & u \frac{\partial}{\partial v} - \bar{v}
\frac{\partial}{\partial \bar{u}} \nonumber \\
L_- & = & v \frac{\partial}{\partial u} - \bar{u}
\frac{\partial}{\partial \bar{v}} \\
L_z \, & = &  \frac{1}{2} \left(-u \frac{\partial}{\partial u} + v
\frac{\partial}{\partial v} + \bar{u} \frac{\partial}{\partial \bar{u}}
- \bar{v} \frac{\partial}{\partial \bar{v}} \right). \nonumber
\end{eqnarray}
These operators acting on the single-particle states $Y_{S,l,m}$ behave
in the expected fashion; that is, $L_+$ raises $m$ by one, $L_-$ lowers
$m$ by one, and $L_z$ has eigenvalue $m$:
\begin{eqnarray}
L_+ Y_{S,l,m} & = & \sqrt{(l-m)(l+m+1)} \, Y_{S,l,m+1} \nonumber \\
L_- Y_{S,l,m} & = & \sqrt{(l+m)(l-m+1)} \, Y_{S,l,m-1} \\
L_z \, Y_{S,l,m} & = & m Y_{S,l,m}. \nonumber
\end{eqnarray}

Another set of operators exists on the sphere when a magnetic monopole
is introduced:
\begin{eqnarray}
T_+ & = & -u \frac{\partial}{\partial \bar{v}} + v
\frac{\partial}{\partial \bar{u}} \nonumber \\
T_- & = & \ \ \: \bar{u} \frac{\partial}{\partial v} - \bar{v}
\frac{\partial}{\partial u} \\
T_z \, & = & \frac{1}{2} \left( u \frac{\partial}{\partial u} + v
\frac{\partial}{\partial v} - \bar{u} \frac{\partial}{\partial \bar{u}}
- \bar{v} \frac{\partial}{\partial \bar{v}} \right). \nonumber
\end{eqnarray}
These operators obey the commutation relations for angular momentum
(SU(2)) just as the $L$ operators do; i.e.\ $[ T_i, T_j ] =
\epsilon_{ijk} T_k$.  They commute with the $L$ operators: $[ \vec{L},
\vec{T} ] = 0$, where $\vec{L} = (L_x, L_y, L_z)$, $\vec{T} = (T_x,
T_y, T_z)$.  Rather than changing the angular momentum quantum number
$m$, however, they change the monopole charge $S$,
\begin{eqnarray}
T_+ Y_{S,l,m} & = & \sqrt{(l-S)(l+S+1)} \, Y_{S+1,l,m} \nonumber \\
T_- Y_{S,l,m} & = & \sqrt{(l+S)(l-S+1)} \, Y_{S-1,l,m} \\
T_z \, Y_{S,l,m} & = & S \, Y_{S,l,m}. \nonumber
\end{eqnarray}
Note that $T_z = \hat{\Omega} \cdot \vec{L}$, where $\hat{\Omega}$ is a
unit vector indicating a position on the sphere.

Because the monopole harmonics $Y_{S,l,m}$ form a complete basis for a
fixed $S$, any wavefunction $\psi$ which describes a particle moving on
the sphere in the presence of a magnetic monopole of charge $S$ must
satisfy
\begin{equation}
T_z \psi = S \psi.
\end{equation}

Now we can calculate the kinetic energy $K$ in (\ref{kinetic_single})
more efficiently using the $T$ operators.  We use the facts that $T^2 =
L^2$ and $[T_+, T_-] = 2 T_z$ to get
\begin{eqnarray}
L^2 = T^2 & = & \frac{1}{2} \left( T_- T_+ + T_+ T_- \right) + T_z^2
\nonumber \\
    & = & T_- T_+ + T_z^2 + T_z \\
	 & = & T_- T_+ + S^2 + S. \nonumber
\end{eqnarray}
Then the kinetic energy operator becomes
\begin{equation}
K = \frac{T_- T_+ + S}{R^2},
\label{excess}
\end{equation}
and $T_- T_+$ acting on a wavefunction $\psi$ is enough to give us the
kinetic energy.  A wavefunction entirely in the LLL has $T_- T_+ \psi_0
= 0$.

These two sets of operators are analogous to the inter-Landau level
operators $a, a^\dagger$ and intra-Landau level operators $b,
b^\dagger$ in the planar geometry.  This is easily seen by comparing
the effects of these operators on the single-particle states in the
plane to the $T_\mp$ and $L_\mp$ operators acting on $Y_{S,l,m}$ on the
sphere.

\subsection{Integer quantum Hall states}
\label{iqheprocedure}

The integer quantum Hall effect (IQHE) wavefunctions are commonly
written as a Slater determinant in which all single-particle states up
to the $\nu^{\rm th}$ Landau level are filled.  Any excitation of this
state must involve raising one or more particles up one Landau level,
producing an energy gap in the spectrum.  This energy gap causes the
many-particle state to be incompressible and creates the quantum Hall
effect.  There is no Landau-level mixing in this wavefunction, since
each Landau level is either completely filled or completely empty.
This wavefunction is strictly valid only at infinite magnetic field and
infinite density.  We introduce Landau-level mixing, and thus find the
dependence of the total energy on the density, by attaching a Jastrow
factor
\begin{equation}
\psi_\alpha = D J_\alpha,
\end{equation}
where $D$ is the Slater determinant.  The Landau-level mixing costs
some kinetic energy but the cost is more than offset by a gain in
potential energy, as correlations in the Jastrow factor allow the
electrons to stay farther apart.

The ground state is a uniform liquid, but excited states will create
regions of charge excess and charge deficit.  In order to create a free
electron and hole in the excited state, we move these regions of charge
excess and deficit as far apart as possible.  On the sphere, this is
best done by keeping the hole near the north pole and the electron near
the south pole. We do this by removing an electron from the
$Y_{S,l,-l}$ state and placing it in the $Y_{S,l+1,l+1}$ state in the
first empty Landau level.  The remaining interaction between hole and
electron is removed by subtracting from the potential energy the
interaction between an electron fixed at the south pole and a hole
fixed at the north pole.

The Monte Carlo algorithm proceeds by moving a single particle, chosen
at random, then calculating the ratio of the new probability density
$|\psi^\prime_\alpha|^2$ to the old $|\psi_\alpha|^2$.  (Primes on a
symbol indicate that it has been computed after the move.) If the new
configuration $\Omega^\prime$ is more likely than the old $\Omega$,
then the move is always accepted; if it is less likely, the move is
accepted with probability
\begin{equation}
\frac{|\psi^\prime_\alpha|^2}{|\psi_\alpha|^2} =
   \frac{{D^\prime}^2}{D^2} \frac{{J_\alpha^\prime}^2}{J_\alpha^2}.
\end{equation}
Because the Jastrow factor is a pair product, the computational cost of
an update after a single move is of order $N$.  At first glance, the
cost of an update to the Slater determinant is also of order $N$, since
we can write
\begin{equation}
D^\prime = \sum_m \phi_m(\Omega^\prime_n) \tilde{D}_{mn},
\end{equation}
where we have moved the $n^{\rm th}$ particle,
$\phi_m(\Omega^\prime_n)$ is the $m^{\rm th}$ single-particle state
after the move, and $\tilde{D}_{mn}$ is the appropriate cofactor.  If
the move is rejected, we need do no more computation.  However, if the
move is accepted, we now have to update all the cofactors with index
$n$.

Fahy {\it et al} \cite{fahy90}, and earlier Ceperley {\it et al}
\cite{ceperley77} have given an algorithm which computes these
cofactors at a cost of order $N^2$.  They begin the calculation by
computing the inverse of the transposed Slater matrix, which gives the
cofactors divided by the determinant
\begin{equation}
\bar{D}_{jk} = \frac{\tilde{D}_{jk}}{D}.
\end{equation}
The ratio of the new determinant to the old is
\begin{equation}
q = \frac{D^\prime}{D} = \sum_m \phi_m(\Omega^\prime_n) \bar{D}_{mn}.
\end{equation}
The cofactors $\tilde{D}_{jk}$ are never calculated explicitly -- they
are kept in memory as the elements of the matrix $\bar{D}_{jk}$.  These
numbers are updated according to
\begin{equation}
\bar{D}^\prime_{jk} = \left\{ \begin{array}{ll}
   \bar{D}_{jk}/q & k = n \\
	\bar{D}_{jk} - \frac{\bar{D}_{jn}}{q} \sum_m \phi_m(\Omega^\prime_n)
\bar{D}_{mk} & k \ne n.
	\end{array} \right.
\end{equation}
Computing the potential energy is now straightforward using
(\ref{coulomb}).

The kinetic energy is computed by applying $\sum_i T^-_i T^+_i$ to
$\psi_\alpha$ and using (\ref{excess}).  Because the $T^\pm_n$ are
first-order differential operators, both the product rule and the chain
rule apply, and we find
\begin{eqnarray}
\frac{\sum_i T^-_i T^+_i \psi_\alpha}{\psi_\alpha} & = &
\frac{1}{DJ_\alpha} \sum_i T^-_i T^+_i(D J_\alpha) \nonumber \\
   & = & \sum_i \left[ \frac{T^-_i J_\alpha}{J_\alpha} \; \frac{T^+_i
D}{D} + \frac{T^-_i T^+_i D}{D} \right. \\
	&   & \mbox{} + \left. \frac{T^-_i D}{D} \; \frac{T^+_i
J_\alpha}{J_\alpha} + \frac{T^-_i T^+_i J_\alpha}{J_\alpha} \right].
\nonumber
\end{eqnarray}
The action of $T^\pm_n$ on the determinant is easily found,
\begin{eqnarray}
\frac{T^\pm_n D}{D} & = & \sum_m T^\pm_n \phi_m(\Omega_n) \bar{D}_{mn}
\label{oneTD} \\
\frac{T^-_n T^+_n D}{D} & = & \sum_m T^-_n T^+_n \phi_m(\Omega_n)
\bar{D}_{mn} \label{twoTD}
\end{eqnarray}
The second quantity (\ref{twoTD}), when summed over all $n$, is just
the excess kinetic energy of the determinant $D$, a constant
independent of the electron positions.  The first quantity
(\ref{oneTD}) can be also be computed quickly, since the
single-particle states are the monopole harmonics
$Y_{S,l,m}(\Omega_n)$, and we have already computed the matrix elements
$\bar{D}_{nm}$.

The action of $T^\pm_n$ on the Jastrow factor is found most
conveniently by defining two rotationally-invariant quantities on the
sphere,
\begin{eqnarray}
r_{ij} & \equiv & u_i v_j - v_i u_j \label{rij} \\
s_{ij} & \equiv & \bar{u}_i u_j + \bar{v}_i v_j. \label{sij}
\end{eqnarray}
Here $|r_{ij}|$ is proportional to the chord distance on the sphere
between particles $i$ and $j$, and $s_{ij}$ is related to the
particle's center of mass.  Their property of rotational invariance is
verified by applying an arbitrary rotation
\begin{equation}
\left( \begin{array}{cc} \bar{u}_0 & \bar{v}_0\\ -v_0 & u_0 \end{array}
\right)
\left( \begin{array}{c} u \\ v \end{array} \right) =
\left( \begin{array}{c} \bar{u}_0 u + \bar{v}_0 v \\ u_0 v - v_0 u
\end{array} \right),
\end{equation}
substituting for $u$ and $v$ in (\ref{rij}) and (\ref{sij}), and using
the identity $u \bar{u} + v \bar{v} = 1$.

The $T^\pm_n$ operators have a particularly simple effect on $r_{ij}$
and $s_{ij}$,
\begin{equation}
\begin{array}{rcl}
T^-_i r_{ij} & = & -s_{ij} \\
T^+_i r_{ij} & = & 0 \nonumber
\end{array}
\hspace{0.5cm}
\begin{array}{rcl}
T^-_i \bar{r}_{ij} & = & 0 \\
T^+_i \bar{r}_{ij} & = & \bar{s}_{ij} \nonumber
\end{array}
\hspace{0.5cm}
\begin{array}{rcl}
T^-_i s_{ij} & = & 0 \\
T^+_i s_{ij} & = & -r_{ij} \nonumber
\end{array}
\hspace{0.5cm}
\begin{array}{rcl}
T^-_i \bar{s}_{ij} & = & \bar{r}_{ij} \\
T^+_i \bar{s}_{ij} & = & 0
\label{trs}
\end{array}
\end{equation}
Other identities are found using $r_{ji} = -r_{ij}$ and $s_{ji} =
\bar{s}_{ij}$.  Then the action of $T^\pm_i$ on the Jastrow factor is
\begin{eqnarray}
\frac{T^\pm_n J_\alpha}{J_\alpha} & = & \frac{1}{J_\alpha} T^\pm_n \exp
\left[ -\alpha \sum_{i<j} c(r_{ij} \bar{r}_{ij}) \right] \\
 & = & -\alpha {\sum_i}^\prime \left[ (T^\pm_n r_{ni})\bar{r}_{ni} +
r_{ni} (T^\pm_n \bar{r}_{ni}) \right] c^\prime(r_{ni} \bar{r}_{ni}),
\nonumber
 \end{eqnarray}
where $c^\prime(x) \equiv \partial c/\partial x$, the prime on the sum
indicates that the sum is over all $i \ne n$, and we have defined a new
pseudopotential function $c(r^2) \equiv u(r)$ equivalent to the old
with a different argument.  Applying (\ref{trs}) gives explicit
expressions.  The quantity $(T^-_n T^+_n J_\alpha)/J_\alpha$ is
evaluated in a similar fashion.

\subsection{Fractional quantum Hall states}

It is well-known that the FQHE occurs at certain fractional filling
factors $\nu$.  In contrast to the IQHE, which is a single-particle
phenomenon, the FQHE is caused by interactions between the particles.
At these fractional fillings the particles can avoid each other best by
staying in high angular momentum states relative to each other.
Excitations in this state must have a gap, since any excitation has to
lower at least one particle's angular momentum with respect to the
others, using a finite amount of energy.

The excitations appear as quasihole-quasielectron pairs, each carrying
a fractional charge.  The energy required to create the pair, the gap
energy $E_g$, determines the magnitude of the cusp that appears in the
ground-state energy with respect to $\nu$, since the states just off
the FQHE filling factors $\nu$ must have a finite number of
quasiparticles.  The magnitude of the cusp is related to the gap
\cite{macdonald86} by
\begin{equation}
E_g = \nu^2 \left[ \left. \frac{\partial E}{\partial \nu} \right|_{\nu
= \nu+} - \left. \frac{\partial E}{\partial \nu} \right|_{\nu = \nu-}
\right]
\label{discontinuity}
\end{equation}

In this paper we will consider only the principal states at $\nu =
1/(p+1), p$ even.  The Laughlin wavefunctions $\psi_L$ describing the
states on the sphere \cite{haldane83} are written $\psi_L = L^{p+1}$
where
\begin{equation}
L \equiv \prod_{i<j} (u_i v_j - v_i u_j).
\end{equation}
We need an explicit wavefunction for the quasihole-quasielectron state
as well.  In order to do so we turn to the composite fermion picture,
where
\begin{equation}
\psi_{CF} = D L^p.
\end{equation}
Here the Slater determinant $D$ is a full Landau level when describing
the principal states.  Since at this filling $D$ is a Van der Monde
determinant, $D=L$ and the two forms are equivalent.

In the composite fermion picture, however, we think of $D$ as
describing the particles, and the remaining factor $L^p$ as attaching
$p$ magnetic flux quanta to each particle.  To complete the picture we
must then project the new wavefunction onto the LLL
\begin{equation}
\psi_{CF}^0 = {\cal P} \left[ D L^p \right],
\end{equation}
where ${\cal P}$ is the projection operator.  To excite a
quasiparticle pair, we follow the IQHE example by moving an electron in
$D$ from the LLL to the second Landau level.  We again want the
quasiparticles as far apart as possible, so we put the quasihole on the
north pole and the quasielectron on the south pole.  As before, we do
this by moving the electron in $Y_{S,S,-S}$ to $Y_{S,S+1,S+1}$.  We
then subtract from the potential energy the interaction of
fractionally-charged particles on opposite poles to model a free
quasihole and quasielectron.

Unfortunately, projecting the composite fermion wavefunction to the LLL
is, to the authors' knowledge, an unsolved problem.  It has been done
for a small (up to 12) number of particles only by explicitly
decomposing the wavefunction in the Slater determinant basis and
keeping only those terms in the LLL \cite{jain}.  There is, however, a
trick recently used by Bonesteel \cite{bonesteel95} which can project a
wavefunction with a single electron, or perhaps two or three, in the
second Landau level to the LLL.  If the occupancy of the second Landau
level is at most one, for example, $\psi_{CF}$ can be decomposed as
\begin{equation}
\psi_{CF} = \psi_{CF}^0 + \psi_{CF}^1,
\end{equation}
where $\psi_{CF}^0$ is entirely in the LLL, and $\psi_{CF}^1$ has
exactly one electron in the second Landau level.  Letting the excess
kinetic energy operator $\sum_n T^-_n T^+_n$ act on $\psi_{CF}$, then
\begin{eqnarray}
& & \sum_n T^-_n T^+_n \psi_{CF} = 0 \; \psi_{CF}^0 + \Delta
\psi_{CF}^1 \nonumber \\
& \Rightarrow & (\sum_n T^-_n T^+_n - \Delta)\psi_{CF} = -\Delta
\psi_{CF}^0
\end{eqnarray}
is all in the LLL.  Here $\Delta = 2S+2$ is the separation between
Landau levels, and we have shifted the zero of the kinetic energy to
ignore zero-point motion.  Remembering the definition in (\ref{rij}),
and normalizing by the unprojected wavefunction, we find
\begin{eqnarray}
\frac{\psi_{CF}^0}{\psi_{CF}} & = & \frac{1}{\psi_{CF}^0 \Delta}
	\left[ \sum_n  T^-_n T^+_n - \Delta \right] \left( D L^p \right)
\nonumber\\
	   & = & \frac{p}{\Delta} \sum_n \frac{T^-_n L}{L} \; \frac{T^+_n
D}{D} + \left( \frac{\Delta_D}{\Delta} - 1 \right)
\end{eqnarray}
where we have used the fact that $T^+_n L = 0$ and
\begin{equation}
\Delta_D D = \sum_n  T^-_n T^+_n D.
\end{equation}
We have already found $(T^+_n D)/D$.  Writing $L = \prod_{i<j} r_{ij}$
and using (\ref{trs}), we find
\begin{equation}
\frac{T^-_n L}{L} = -{\sum_i}^\prime \frac{s_{ni}}{r_{ni}}.
\label{tl}
\end{equation}

In order to introduce Landau-level mixing into the wavefunction, we
again attach a Jastrow factor
\begin{equation}
\psi_\alpha = {\cal P} \left[ D L^p \right] J_\alpha.
\end{equation}
The potential energy, as before, can be found by averaging
(\ref{coulomb}) over all configurations generated in the Monte Carlo
code by $|\psi_\alpha|^2$.

The kinetic energy for the ground state can be found in a
straightforward manner, since the ground-state variational wavefunction
is
\begin{equation}
\psi_\alpha = L^{p+1} J_\alpha.
\end{equation}
To calculate the kinetic energy, we use
\begin{eqnarray}
\frac{T^-_n T^+_n \psi_\alpha}{\psi_\alpha} & = & \frac{1}{L^{p+1}
J_\alpha}
	\sum_n T^-_n T^+_n (L^{p+1} J_\alpha) \nonumber \\
& = & (p+1) \sum_n \frac{T^-_n L}{L} \; \frac{T^+_n J_\alpha}{J_\alpha}
+
	\sum_n \frac{T^-_n T^+_n J_\alpha}{J_\alpha},
\end{eqnarray}
where we have used the fact that $T^+_n L = 0$.  We have previously
shown how to compute each of these terms.

Computing the kinetic energy for the quasihole-quasielectron state is
more involved, due to the presence of the projection operator
${\cal P}$.  The complexity arises from the fact that our
implementation of ${\cal P}$ involves a product of two $T$
operators, and the kinetic energy uses another product of two $T$
operators.  We need to calculate the quantity
\begin{equation}
\frac{\sum_m T^-_m T^+_m (\psi_{CF}^0 J_\alpha)}{\psi_{CF} J_\alpha} =
	\sum_m \frac{T^-_m\psi_{CF}^0}{\psi_{CF}} \; \frac{T^+_m
J_\alpha}{J_\alpha}
  + \frac{\psi_{CF}^0}{\psi_{CF}} \sum_m \frac{T^-_m T^+_m
J_\alpha}{J_\alpha}.
\end{equation}
The only term we have not yet found is $(T^-_m\psi_{CF}^0)/\psi_{CF}$.
We have
\begin{eqnarray}
\frac{T^-_m\psi_{CF}^0}{\psi_{CF}} & = & \frac{T^-_m{\cal
P}[DL^p]}{DL^p} \nonumber \\
	& = & \frac{1}{DL^p} T^-_m \left\{ \left[ \frac{p}{\Delta} \sum_n
\frac{T^-_n L}{L} \; \frac{T^+_n D}{D} + \left( \frac{\Delta_D}{\Delta}
- 1 \right) \right] DL^p \right\} \nonumber \\
		& = & \frac{p(p-1)}{\Delta} \frac{T^-_m L}{L} \sum_n \frac{T^-_n
L}{L} \; \frac{T^+_n D}{D} + \mbox{} \\
		& & \frac{p}{\Delta} \sum_n \left[ \frac{T^-_m T^-_n L}{L} \;
\frac{T^+_n D}{D} + \frac{T^-_n L}{L} \; \frac{T^-_m T^+_n D}{D}
\right] + \mbox{} \nonumber \\
		& & \left( \frac{\Delta_D}{\Delta} - 1 \right) \left[ \frac{T^+_m
D}{D} + p \frac{T^-_m L}{L} \right]. \nonumber
\end{eqnarray}
The two terms not yet computed are $(T^-_m T^-_n L)/L$ and $(T^-_m
T^+_n D)/D$.

We can find $(T^-_m T^-_n L)/L$ in a straightforward manner, by
applying (\ref{trs}) and (\ref{tl}). First we note that
\begin{equation}
T^-_m \left( \frac{s_{ni}}{r_{ni}} \right) = \delta_{mn} \left(
\frac{s_{ni}}{r_{ni}} \right)^2 - (1-\delta_{mn})
\frac{\delta_{mi}}{r_{ni}^2},
\end{equation}
where we assume that $n \ne i$ and we use the identity $s_{ij}
\bar{s}_{ij} + r_{ij} \bar{r}_{ij} = 1$.  Then we have
\begin{eqnarray}
\frac{T^-_m T^-_n L}{L} & = & \left( \sum_{i, i\ne n}
\frac{s_{ni}}{r_{ni}} \right) \left( \sum_{i, i \ne m}
\frac{s_{mi}}{r_{mi}} \right) + \mbox{} \\
& & \frac{1-\delta_{mn}}{r_{mn}^2} - \delta_{mn} \sum_{i,i \ne m}
\left( \frac{s_{mi}}{r_{mi}} \right)^2. \nonumber
\end{eqnarray}
The first two sums have been computed earlier, and the sum in the last
term enters only when $n = m$.

The second term $(T^-_m T^+_n D)/D$ is computed easily when $m=n$,
\begin{equation}
\frac{T^-_n T^+_n D}{D} = \sum_i T^-_n T^+_n \phi_i(\Omega_n)
\bar{D}_{in}.
\end{equation}
When $m \ne n$ we use the identity (derived in the appendix)
\begin{eqnarray}
\frac{T^-_m T^+_n D}{D} & = & \left( \sum_i T^+_n \phi_i(\Omega_n)
\bar{D}_{in} \right) \left( \sum_i T^-_m \phi_i(\Omega_m) \bar{D}_{im}
\right) - \mbox{} \label{ttd} \\
& & \left( \sum_i T^+_n \phi_i(\Omega_n) \bar{D}_{im} \right) \left(
\sum_i T^-_m \phi_i(\Omega_m) \bar{D}_{in} \right), \nonumber
\end{eqnarray}
where the $\phi_i(\Omega_j)$ are again the single-particle
wavefunctions.  The $N^3$ scaling of the time required to compute the
kinetic energy of $\psi_{CF}^0$ arises from (\ref{ttd}), since
(\ref{ttd}) has to be computed for all $N^2$ pairs $(m,n)$ and each
computation scales as $N$.

\section{Implementation of Monte Carlo method}

The implementation of the Monte Carlo procedure described above
requires some adjustment in practice.  The pseudopotential $u(r)$,
(\ref{pseudopotential}), if put directly on the sphere, will cause wild
fluctuations in the kinetic energy calculation.  The cause is found in
the cusp at $r=2R$, introduced in the pseudopotential by the mapping to
the sphere.  In effect, a $\delta$-function in the second derivative of
the trial wavefunction appears if $\partial u/\partial r \ne 0$ at
$r=2R$.  We remove the cusp by making a linear combination of $u(r)$'s
with different arguments,
\begin{equation}
\tilde{u}(r) = u(r) + u(4R-r).
\end{equation}
Then
\begin{equation}
\left. \frac{\partial \tilde{u}}{\partial r} \right|_{r=2R} =
\left. \frac{\partial u}{\partial r} \right|_{r=2R} -
\left. \frac{\partial u}{\partial r} \right|_{4R-r = 2R},
\end{equation}
$\partial u/\partial r = 0$ at $r=2R$, and the cusp is removed.  The
Jastrow factor is now
\begin{equation}
J_\alpha = \exp \left[ - \alpha \sum_{i<j} \tilde{u}(|r_{ij}|) \right],
\end{equation}
where we have used the definition (\ref{rij}).  With the variational
parameter $\alpha$ appearing only outside the pseudopotential, we can
compute potential and kinetic energies for a range of $\alpha$, then do
the minimization with respect to $\alpha$ for any number of values for
$r_s$ without repeating the Monte Carlo calculation.

Calculating the IQHE ground-state potential and kinetic energies by the
Monte Carlo procedure in section \ref{iqheprocedure} is
straightforward, since the trial wavefunction is simple.  The variance
in the result is small and scales as $\sqrt{N}$.  The energy gap to the
first excited state is more problematic, however.  If we move an
electron up to the next Landau level, creating an electron-hole pair,
we must evaluate the pair energy by subtracting the ground state total
energy from the excited state total energy.  The variance in the Monte
Carlo results for either state is then of the same order as the energy
gap, and the results of the calculation are obscured by the statistical
noise.  Increasing $N$ only worsens the problem, since the energy gap
remains approximately constant with $N$, but the variance in the
energies gets larger.

We circumvent this difficulty in the manner recommended by Ceperley in
\cite{binderbook}, by sampling from the ground-state wavefunction and
computing the excited state energy and ground state energy
simultaneously.  The two energies are then correlated: an upward
fluctuation in the ground state energy always correlates with a nearly
equivalent upward fluctuation in the excited state energy, for example,
and the variance of the difference is small.

The computation of the excited state energy is implemented in the
following way:  the ground state energy is calculated in the Monte
Carlo code as
\begin{eqnarray}
\left< \alpha | H | \alpha \right>_{\rm gnd} & = & \int \! d\Omega
\left| \psi_\alpha^{\rm gnd} \right|^2  \frac{H \psi_\alpha^{\rm
gnd}}{\psi_\alpha^{\rm gnd}} \nonumber \\
 & = & \sum_n \frac{H \psi_\alpha^{\rm gnd}}{\psi_\alpha^{\rm gnd}},
\label{simplemontecarlo}
\end{eqnarray}
since the sample configurations are drawn from $\psi_\alpha^{\rm
gnd}(\Omega)$.  (As before, we define $\Omega = \{\Omega_1, \cdots,
\Omega_N\}$, where $\Omega_i$ is the position of the $i^{\rm th}$
electron.)  We write the excited state energy as
\begin{eqnarray}
\left< \alpha | H | \alpha \right>_{\rm ex} & = & \int \! d\Omega
\left| \psi_\alpha^{\rm ex} \right|^2  \frac{H \psi_\alpha^{\rm
ex}}{\psi_\alpha^{\rm ex}} \nonumber \\
& = & \kappa \int \! d\Omega \left| \psi_\alpha^{\rm gnd} \right|^2
\left| \frac{\psi_\alpha^{\rm ex}}{\psi_\alpha^{\rm gnd}} \right|^2
\frac{H \psi_\alpha^{\rm ex}}{\psi_\alpha^{\rm ex}} \nonumber \\
& = & \kappa \sum_{n} \left| \frac{\psi_\alpha^{\rm
ex}(\Omega_n)}{\psi_\alpha^{\rm gnd}(\Omega_n)} \right|^2  \frac{H
\psi_\alpha^{\rm ex}(\Omega_n)}{\psi_\alpha^{\rm ex}(\Omega_n)},
\label{messymontecarlo}
\end{eqnarray}
where the normalization $\kappa$ is no longer 1,
\begin{eqnarray}
\kappa^{-1} & = & \int \! d\Omega \left| \psi_\alpha^{\rm gnd}
\right|^2 \left| \frac{\psi_\alpha^{\rm ex}}{\psi_\alpha^{\rm gnd}}
\right|^2 \nonumber \\
& = & \sum_n \left| \frac{\psi_\alpha^{\rm
ex}(\Omega_n)}{\psi_\alpha^{\rm gnd}(\Omega_n)} \right|^2.
\label{normalization}
\end{eqnarray}

Calculating the ground state energy for the FQHE states is
comparatively straightforward as well, using (\ref{simplemontecarlo}).
The FQHE ground state energies in this paper were all done in this
manner.  Because of the complexity of the excited state wavefunction,
however, it is difficult to follow the recipe in
(\ref{messymontecarlo}).  Instead, we sample from the unprojected
wavefunction
\begin{equation}
\psi_\alpha^{\rm un} = DL^pJ_\alpha.
\end{equation}
We compute both excited state and ground state wavefunctions similarly
to (\ref{messymontecarlo}).  Define $\psi_\alpha^{\rm gnd} = {\cal
P} \left[ D_{\rm gnd}L^p \right] J_\alpha = L^{p+1} J_\alpha$ and
$\psi_\alpha^{\rm ex} = {\cal P} \left[ D_{\rm ex}L^p \right]
J_\alpha$.  For the ground state,
\begin{eqnarray}
\left< \alpha | H | \alpha \right>_{\rm gnd} & = & \int \! d\Omega
\left| \psi_\alpha^{\rm gnd} \right|^2  \frac{H \psi_\alpha^{\rm
gnd}}{\psi_\alpha^{\rm gnd}} \nonumber \\
& = & \kappa_{\rm gnd} \int \! d\Omega \left| \psi_\alpha^{\rm un}
\right|^2 \left| \frac{\psi_\alpha^{\rm gnd}}{\psi_\alpha^{\rm un}}
\right|^2  \frac{H \psi_\alpha^{\rm gnd}}{\psi_\alpha^{\rm gnd}}
\nonumber \\
& = & \kappa_{\rm gnd} \sum_{n} \left| \frac{\psi_\alpha^{\rm
gnd}(\Omega_n)}{\psi_\alpha^{\rm un}(\Omega_n)} \right|^2  \frac{H
\psi_\alpha^{\rm gnd}(\Omega_n)}{\psi_\alpha^{\rm gnd}(\Omega_n)},
\label{checkmontecarlo}
\end{eqnarray}
and for the excited state
\begin{eqnarray}
\left< \alpha | H | \alpha \right>_{\rm ex} & = & \int \! d\Omega
\left| \psi_\alpha^{\rm ex} \right|^2  \frac{H \psi_\alpha^{\rm
ex}}{\psi_\alpha^{\rm ex}} \nonumber \\
& = & \kappa_{\rm ex} \int \! d\Omega \left| \psi_\alpha^{\rm un}
\right|^2 \left| \frac{\psi_\alpha^{\rm ex}}{\psi_\alpha^{\rm un}}
\right|^2  \frac{H \psi_\alpha^{\rm ex}}{\psi_\alpha^{\rm ex}}
\nonumber \\
& = & \kappa_{\rm ex} \sum_{n} \left| \frac{\psi_\alpha^{\rm
ex}(\Omega_n)}{\psi_\alpha^{\rm un}(\Omega_n)} \right|^2  \frac{H
\psi_\alpha^{\rm ex}(\Omega_n)}{\psi_\alpha^{\rm ex}(\Omega_n)}.
\end{eqnarray}
The normalizations $\kappa_{\rm gnd}$ and $\kappa_{\rm ex}$ are
calculated in the same way as (\ref{normalization}).

We checked the results of this calculation by comparing the ground
state energies found using (\ref{checkmontecarlo}) to the ground state
energies found using (\ref{simplemontecarlo}).  The energies were
identical to within the statistical variations.

The initial electron configuration for each Monte Carlo run must be
chosen carefully to avoid underflow of the computer's arithmetic.  The
probability of generating a configuration from the trial wavefunction
identical to that made by placing electrons at random on the sphere is
so low that the initial calculation of the Slater determinant $D$ will
invariably underflow.  To work around this problem we use an initial
configuration which keeps the particles as far apart as possible.  The
probability of drawing this configuration from the trial wavefunction
is much higher than that of a random configuration, so the initial
computation of $D$ will not underflow.

This configuration is generated using some code borrowed from Jon Leech
\cite{leech89}.  It begins with six electrons; two on the north and
south poles and four evenly distributed around the equator, forming the
vertices of an octahedron.  Each face is a triangle, which can be
subdivided into four smaller triangles, as in Figure \ref{triangles}.
Placing electrons at the vertices of these new triangles generates a
configuration of 18 electrons.  Repeating the process generates another
configuration of 66 electrons, and so on.  The electrons are, of
course, not directly on the vertices of the new triangles, but lie on
the surface of the sphere, on a line joining each triangle vertex to
the origin.  A few electrons can be removed from this configuration, as
needed, without underflowing the computer arithmetic.

\section{Results}

We have obtained exchange-correlation energies $\bar{\varepsilon}_{\rm
xc}$ and gap energies $\Delta_{\rm eh}$ at $\nu = 1/7$, 1/5, 1/3, 1,
and 2 as a function of $r_s$, using a model \cite{zhang86} of the
electron-electron interaction
\begin{equation}
V(r) = \frac{1}{\sqrt{r^2+\lambda^2}}.
\label{soft_interaction}
\end{equation}
Setting $\lambda = 0$ recovers the Coulomb interaction.  The softer
interactions modeled by (\ref{soft_interaction}) more closely
approximate the interaction experienced by electrons in a GaAs-AlGaAs
heterojunction.  We have calculated $\bar{\varepsilon}_{\rm xc}$ and
$\Delta_{\rm eh}$ for $\lambda = 0$, 0.2, 0.5, and 1.0.

Numerical fits for our computed exchange-correlation energies are given
in Tables \ref{coulomb_energy}-\ref{lambda10_energy}.  The Monte Carlo
results are modeled using an order (2,2) rational function fit,
\begin{equation}
\bar{\varepsilon}_{\rm xc} = \frac{a_0 + a_1 r_s + a_2 r_s^2}{1 + b_1
r_s + b_2 r_s^2}.
\end{equation}
This form ensures that the energy $\bar{\varepsilon}_{\rm xc}$ remains
finite as $r_s \rightarrow \infty$.  The Monte Carlo code produced good
results from $r_s = 0$ to $r_s = 50$ in most cases, although when
$\bar{\varepsilon}_{\rm xc}$ changed rapidly, the code gave results
only up to $r_s = 35$.  This was true only for the Coulomb interaction
at $\nu = 1$ and 2, with fitting parameters given in Table
\ref{coulomb_energy}.  A typical set of results is shown in Figure
\ref{coulomb_graph}.

The uncertainty in the numbers from the Monte Carlo code are given in
the tables by showing the standard deviation in the values of the last
digit in the parameter $a_0$ in parentheses.  The errors in
$\bar{\varepsilon}_{\rm xc}$ for higher $\lambda$ are much smaller than
for the Coulomb case, since for a given change in inter-electron
spacing, the softer interactions produce a smaller (sometimes much
smaller) change in the total energy of the system.  The fluctuations in
the instantaneous energy in the Monte Carlo code are then smaller, and
the corresponding results more accurate.

Numerical fits for the gap energies $\Delta_{\rm eh}$ are given in
Tables \ref{coulomb_gap}-\ref{lambda10_gap}.  Here we also use a
rational function fit, but since the uncertainty in the gap energies
are relatively large we use only an order (1,1) fit,
\begin{equation}
\Delta_{\rm eh} = \frac{a_0 + a_1 r_s}{1 + b_1 r_s}.
\end{equation}
Gap energies for $\nu = 1/3$ and 1/7 are shown in Figures
\ref{gap3_graph} and \ref{gap7_graph}, respectively.

These energies were calculated with an 18-particle system, rather than
the 66-particle system used to calculate the exchange-correlation
energies.  Fluctuations in the larger system tend to overwhelm the gap
energy, since only one particle (or composite particle) is involved in
creating the excitation whose energy we are measuring.

The quasihole-quasielectron (or in the case of integer $\nu$,
hole-electron) interaction between particles of opposite charge on the
north and south poles is removed by subtracting the interaction of
point charges of appropriate magnitude at the same places.  The
uncertainty in the results is given, as before, in the numbers in
parentheses beside the values for $a_0$ in Tables
\ref{coulomb_gap}-\ref{lambda10_gap}.  We keep three digits in the
fitting parameters other than $a_0$ to remove errors in the fit caused
by truncation.

\section{Summary}

We have calculated exchange-correlation energies and energy gaps
between ground state and first excited state for quantum Hall states at
various filling factors $\nu$ using a model electron-electron
interaction which is realistic for electrons in GaAs-AlGaAs
heterostructures.  Including Landau-level mixing in the calculation
allows us to present results at finite magnetic field and electron
density.  Our results are presented in the form of numerical fits to
simple rational functions, convenient for use in current-density
functional theory calculations.  Calculating energy gaps as well as
ground state energies allows us to model the energy near the quantum
Hall state realistically.  This work has been supported by the ONR and 
the NSF.

\appendix
\section{Two single-particle operators acting on a Slater determinant}

Let the two single-particle operators be $\hat{O}_i$ and $\hat{O}_j$,
where $i$ and $j$ are particle indices and $i \ne j$.  We wish to
calculate
$(\hat{O}_i \hat{O}_j D)/D$ where $D$ is a Slater determinant composed
of single-particle states $\phi_i(\Omega_j)$.  In the following
$\tilde{D}_{ij}$ will be the $i,j^{\rm th}$ cofactor of the Slater
matrix.  Transposing and taking the inverse of the Slater matrix, we
find $\bar{D}_{ij} = \tilde{D}_{ij}/D$.  The first operator acting on
$D$ gives
\begin{equation}
\hat{O}_j D = \sum_k \hat{O}_j \phi_k(\Omega_j) \tilde{D}_{kj}.
\label{dtp}
\end{equation}
The new object $\hat{O}_j D$ can be thought of as a new Slater
determinant $D^\prime$.  The new Slater matrix is identical to the old
Slater matrix with the exception of the $j^{\rm th}$ column, which
changes from $\phi_m(\Omega_j)$ to $\hat{O}_j \phi_m(\Omega_j), 1 \!
\le \! m \! \le \! N$.
Given the old inverse matrix elements $\bar{D}_{ij}$, we would like to
find the new elements $\bar{D}^\prime_{ij}$.  The Sherman-Morrison
formula (equations (2.7.4) and (2.7.5) of \cite{numrecipes}) gives
\begin{equation}
\bar{D}^\prime_{ki} = \bar{D}_{ki} - \frac{z_i w_k}{1+z_j}
\label{ninv}
\end{equation}
where we define (following the notation in \cite{numrecipes})
\begin{eqnarray}
u_k & = & \hat{O}_j \phi_k(\Omega_j) - \phi_k(\Omega_j) \nonumber \\
v_k & = & \delta_{jk} \nonumber \\
z_k & = & \sum_m \bar{D}_{mk} u_m \label{nrvars} \\
w_k & = & \sum_m \bar{D}_{km} v_m = \bar{D}_{kj}. \nonumber
\end{eqnarray}
Substituting, we find that
\begin{eqnarray}
z_j & = & \sum_k \hat{O}_j \phi_k(\Omega_j) \bar{D}_{kj} - \sum_k
\phi_k(\Omega_j) \bar{D}_{kj} \nonumber \\
& = & \frac{\hat{O}_j D}{D} - 1.
\end{eqnarray}
Now let the second operator $\hat{O}_i$ act on the new Slater
determinant
\begin{eqnarray}
\frac{\hat{O}_i D^\prime}{D} & = & \frac{1}{D} \sum_k \hat{O}_i
\phi_k(\Omega_i) \left( \bar{D}_{ki}^\prime D^\prime \right) \nonumber
\\
& = & \frac{D^\prime}{D} \left[ \sum_k \hat{O}_i \phi_k(\Omega_i)
\bar{D}_{ki} - \sum_k \hat{O}_i \phi_k(\Omega_i) z_i \bar{D}_{kj}
\left( \frac{\hat{O}_j D}{D} \right)^{-1} \right], \label{odd}
\end{eqnarray}
where we have used (\ref{ninv}) and (\ref{nrvars}).
Recognizing that $D^\prime/D = (\hat{O}_j D)/D$ and using (\ref{dtp}),
we rewrite (\ref{odd}) as
\begin{equation}
\frac{\hat{O}_i \hat{O}_j D}{D} = \frac{\hat{O}_j D}{D} \;
\frac{\hat{O}_i D}{D} - z_i \sum_k \hat{O}_i \phi_k(\Omega_i)
\bar{D}_{kj}.
\end{equation}
Here, using (\ref{nrvars}) and the fact that $\bar{D}_{ki}$ is an
inverse matrix element
\begin{eqnarray}
z_i & = & \sum_k \bar{D}_{ki} \left[ \hat{O}_j \phi_k(\Omega_j) -
\phi_k(\Omega_j) \right] \nonumber \\
& = & \sum_k \hat{O}_j \phi_k(\Omega_j) \bar{D}_{ki} - \delta_{ij},
\end{eqnarray}
but $i \ne j$ by assumption, so we obtain the result,
\begin{eqnarray}
\frac{\hat{O}_i \hat{O}_j D}{D} & = & \left( \sum_k \hat{O}_j
\phi_k(\Omega_j) \bar{D}_{kj} \right) \left( \sum_k \hat{O}_i
\phi_k(\Omega_i) \bar{D}_{ki} \right) - \mbox{} \\
& & \left( \sum_k \hat{O}_j \phi_k(\Omega_j) \bar{D}_{ki} \right)
\left( \sum_k \hat{O}_i \phi_k(\Omega_i) \bar{D}_{kj} \right).
\nonumber
\end{eqnarray}

\section{Spline interpolation between QHE states}

Interpolation between energies $E_j$ of adjacent quantum Hall states in
filling factor $\nu$ can be conveniently accomplished using cubic
splines.  The energy cusp at each state is included in the
interpolation by requiring that the discontinuity in derivative of
energy $E$ be given by (\ref{discontinuity}), rather than requiring
that it be zero, as normally done.  If the filling factor (energy) of
the $j^{\rm th}$ state is given as $\nu_j$  ($E_j$), and the
discontinuities in derivative are $\Delta_j$, the equations defining
the spline parameters $E_j^{\prime \prime}$, following (3.3.7) of
\cite{numrecipes}, are
\begin{equation}
\frac{\nu_j - \nu_{j-1}}{6} E_{j-1}^{\prime \prime} +
\frac{\nu_{j+1} - \nu_{j-1}}{3} E_j^{\prime \prime} +
\frac{\nu_{j+1} - \nu_j}{6} E_{j+1}^{\prime \prime} =
\frac{E_{j+1} - E_j}{\nu_{j+1} - \nu_j} -
\frac{E_j - E_{j-1}}{\nu_j - \nu_{j-1}} + \Delta_j
\end{equation}
Solving this system of equations for the $E_j^{\prime \prime}$
parameters, and using them following \cite{numrecipes} section 3.3, the
interpolation in energy between states at a given $r_s$ is easily
found.  Figure \ref{spline_graph} is an example, showing the cusps at
$r_s = 0$.  Of course, this interpolation ignores the heirarchy states
completely.

\begin{figure}
\caption{Pseudopotential $u(r)$ for various $\alpha$ with cusp
condition satisfied (\protect\ref{pseudopotential}) and not satisfied
(\protect\ref{oldpseudo}).  The solid lines correspond to
(\protect\ref{pseudopotential}) and the dashed lines to
(\protect\ref{oldpseudo}).
% ; i.e.\ $u_0(r) = \alpha/\sqrt{r}$.
\label{pseudopotential_graph}}
\end{figure}

\epsffile{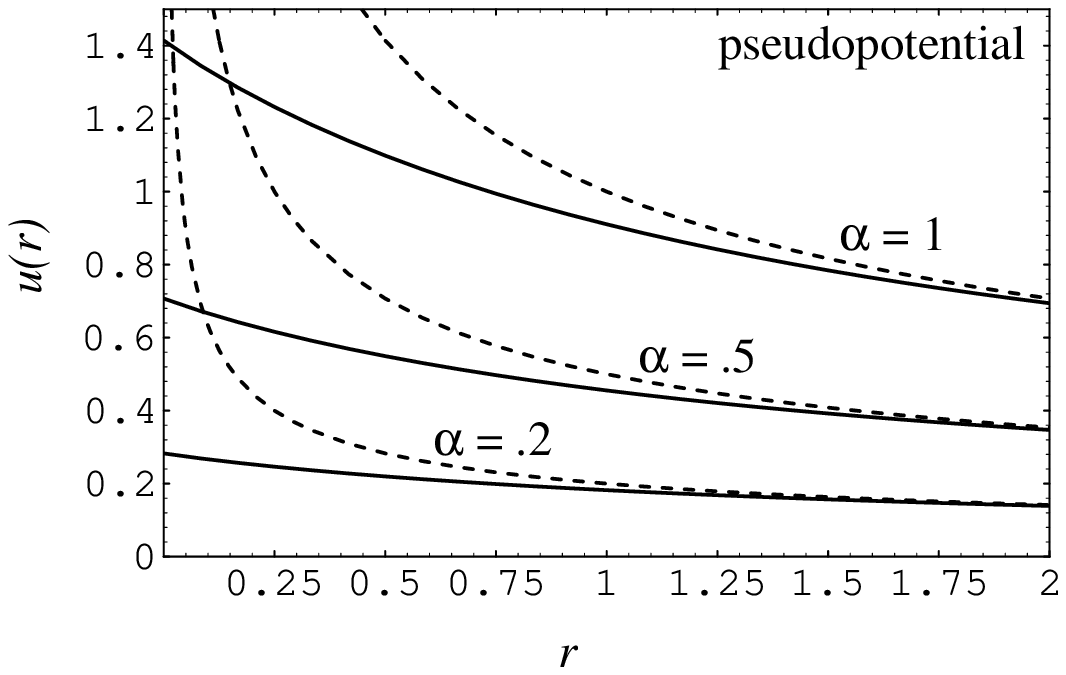}

\newpage
\begin{figure}
\caption{Subdivision of a triangular octahedron face into smaller
triangles as part of the initialization procedure for the Monte Carlo
code.
\label{triangles}}
\end{figure}

\epsffile{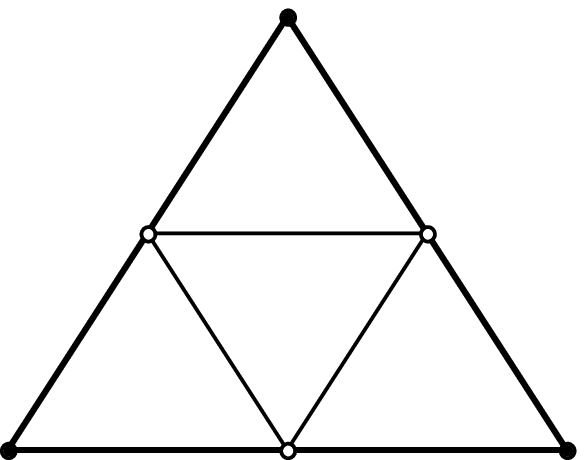}

\newpage
\begin{figure}
\caption{Correlation energy $\bar\varepsilon_{\rm xc}$ as a function of
$r_s$ at various filling factors $\nu$.  Landau-level mixing is more
pronounced at higher filling factors, as expected.
\label{coulomb_graph}}
\end{figure}

\epsffile{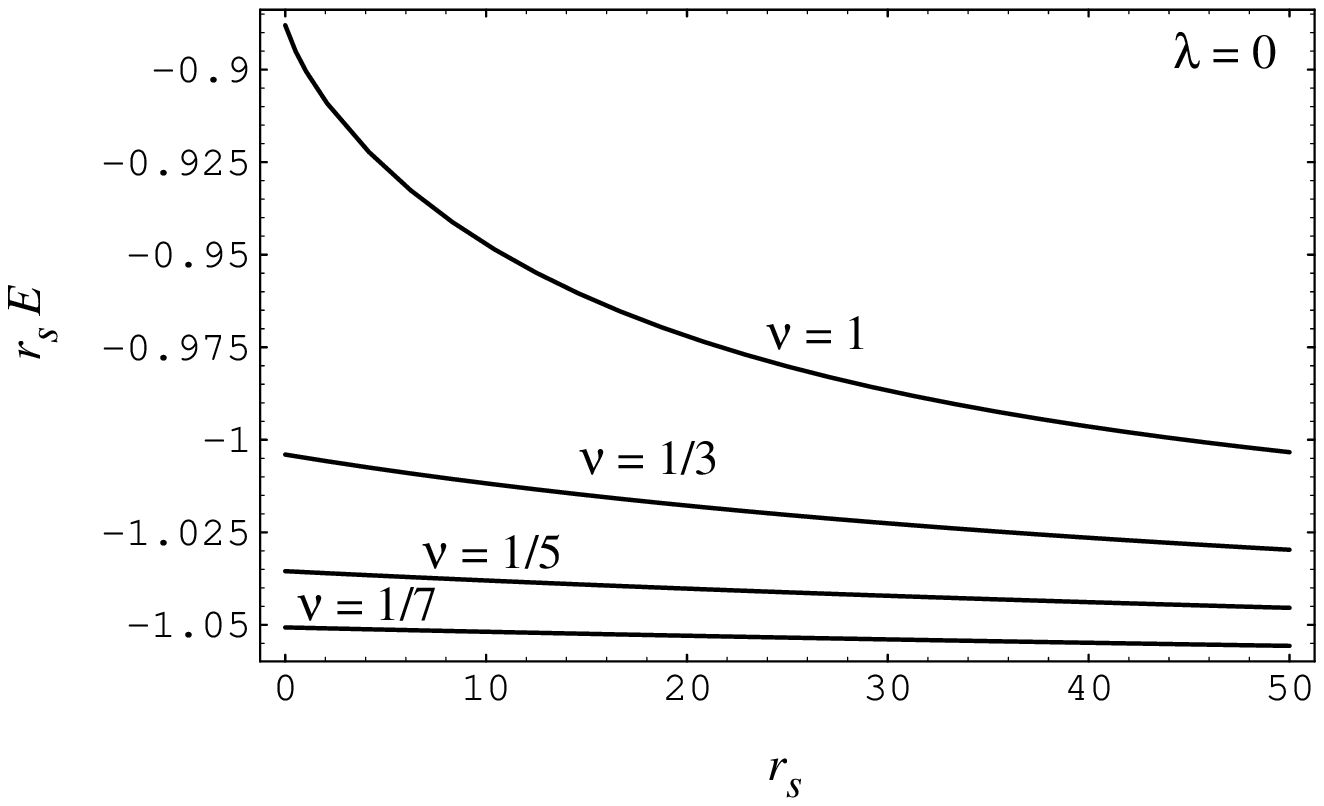}

\newpage
\begin{figure}
\caption{Quasiparticle gap $\Delta_{\rm eh}$ at $\nu = 1/3$ as a
function of $r_s$.
\label{gap3_graph}}
\end{figure}

\epsffile{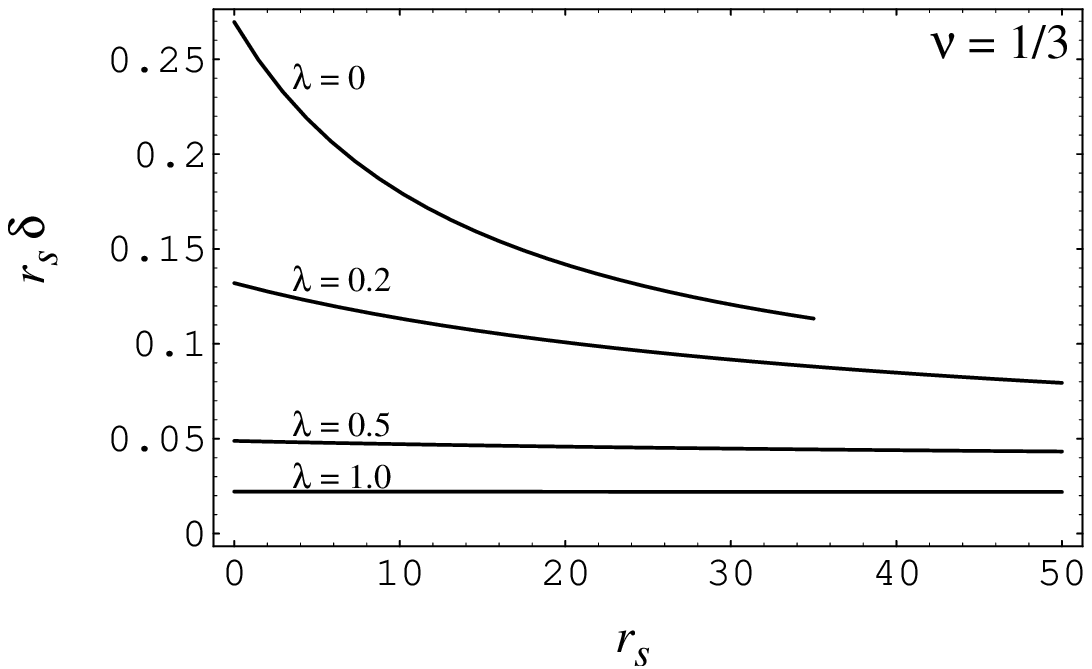}

\newpage
\begin{figure}
\caption{Quasiparticle gap $\Delta_{\rm eh}$ at $\nu = 1/7$ as a
function of $r_s$.  Landau-level mixing continues to have an effect
here, although the effect on total energy is miniscule.
\label{gap7_graph}}
\end{figure}

\epsffile{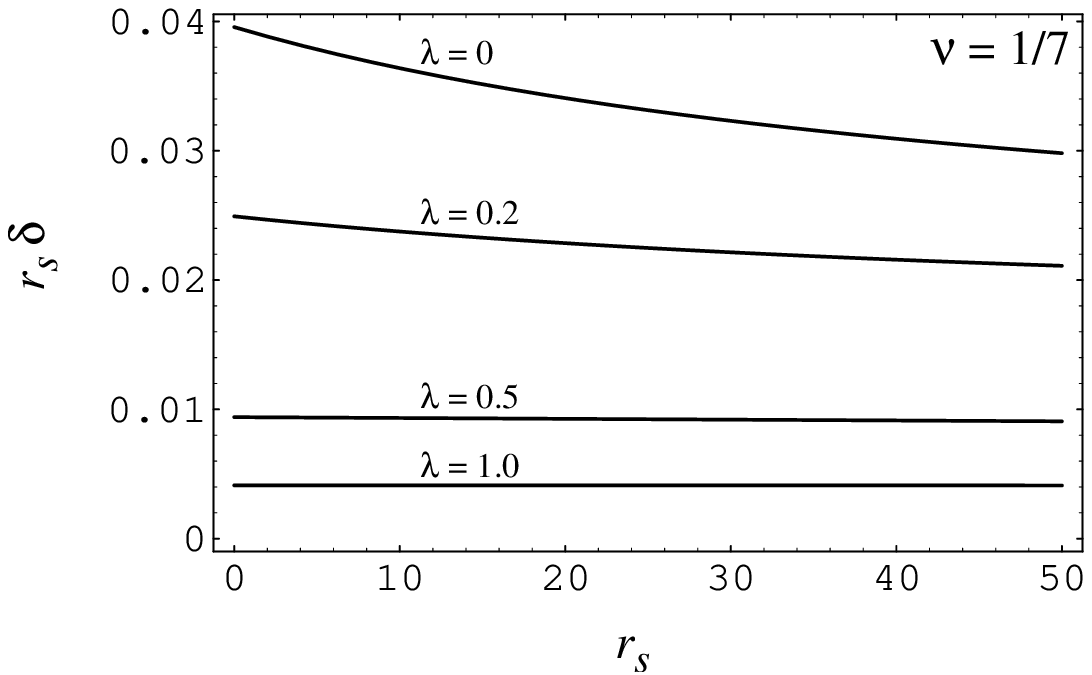}

\newpage
\begin{figure}
\caption{Spline interpolation between energies at $\nu = 1/3$, 1/5, and
1/7, showing cusps in energy at principal FQHE states.  Here $r_s$ = 0.
 The dotted line is a smooth cubic spline interpolation between FQHE
states.
\label{spline_graph}}
\end{figure}

\epsffile{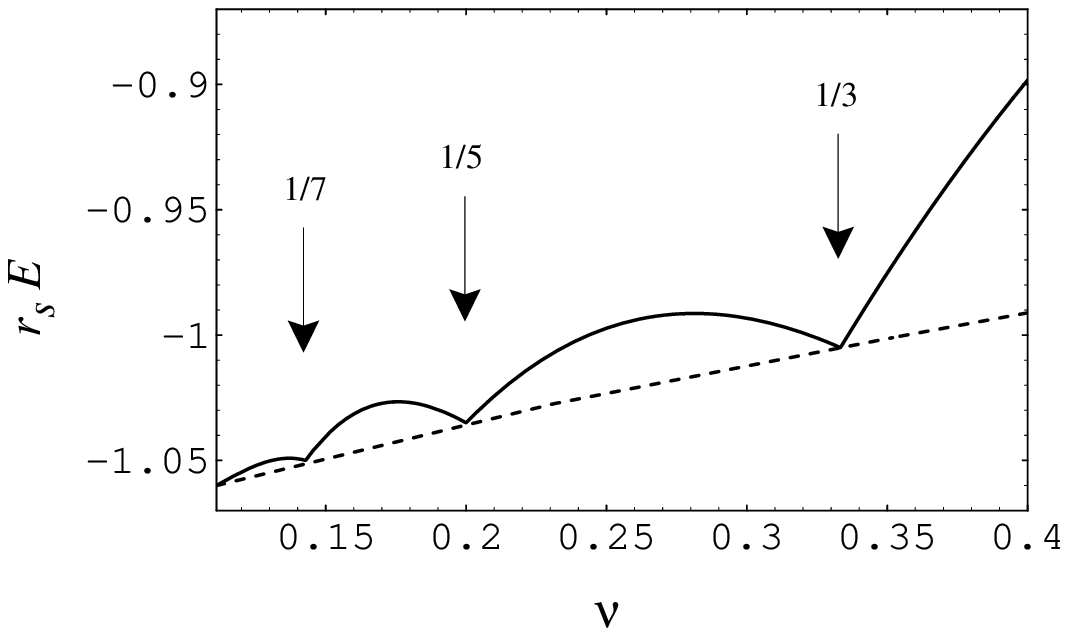}

\begin{table}
\caption[]{Parameters for a (2,2) rational function fit to the
correlation energy $\bar{\varepsilon}_{xc}$ for the Coulomb interaction
($\lambda = 0$).}
\label{coulomb_energy}
\end{table}

\begin{tabular}{|c|clllll|} \hline
\makebox[1.5cm]{$\nu$} & \makebox[0.5cm]{} & \makebox[2cm]{$a_0$} &
\makebox[2cm]{$a_1$} & \makebox[2cm]{$b_1$} & \makebox[2cm]{$a_2 \times
10^3$} & \makebox[2cm]{$b_2 \times 10^3$} \\ \hline
2    & & -0.864(2) & -0.3883   & 0.4221    & -19.85 & 18.97 \\
1    & & -0.888(2)  & -0.7518   & 0.8281    & -41.28 & 39.52 \\
1/3  & & -1.004(1)  & -0.01561  & 0.01466   &   &  \\
1/5  & & -1.0355(8) & -0.007768 & 0.007241  &   &  \\
1/7  & & -1.0507(5) & -0.004958 & 0.004602  &   &  \\ \hline
\end{tabular}

\begin{table}
\caption[]{Parameters for a (2,2) rational function fit to the
correlation energy $\bar{\varepsilon}_{xc}$ with $\lambda = 0.2$.}
\label{lambda02_energy}
\end{table}

\begin{tabular}{|c|clllll|} \hline
\makebox[1.5cm]{$\nu$} & \makebox[0.5cm]{} & \makebox[2.2cm]{$a_0$} &
\makebox[2.2cm]{$a_1$} & \makebox[2.2cm]{$b_1$} & \makebox[2.2cm]{$a_2
\times 10^3$} & \makebox[2.2cm]{$b_2 \times 10^3$} \\ \hline
2    & & -1.7220(1)   & -0.2843    & 0.4221    & -4.081 & 2.327 \\
1    & & -1.7322(1)  & -0.093231  & 0.053603  & -0.21638 & 0.12166 \\
1/3  & & -1.75085(5) & -0.00677   &   &   &  \\
1/5  & & -1.75623(5) & -$8.2 \times 10^{-6}$ &   &   &  \\
1/7  & & -1.75891(5) & -$3.9 \times 10^{-6}$ &   &   &  \\ \hline
\end{tabular}

\newpage

\begin{table}
\caption[]{Parameters for a (1,1) rational function fit to the
correlation energy $\bar{\varepsilon}_{xc}$ with $\lambda = 0.5$.}
\label{lambda05_energy}
\end{table}

\begin{tabular}{|c|clll|} \hline
\makebox[1.5cm]{$\nu$} & \makebox[0.5cm]{} & \makebox[2.5cm]{$a_0$} &
\makebox[2.5cm]{$a_1$} & \makebox[2.5cm]{$b_1$} \\ \hline
2    & & -3.22068(1)   & -0.0460183    & 0.0142792 \\
1    & & -3.22281(1) & -$4.458 \times 10^{-6}$ &  \\
1/3  & & -3.22492(1) & -$2.133 \times 10^{-7}$ & \\
1/5  & & -3.22540(1) & -$5.4 \times 10^{-8}$ & \\
1/7  & & -3.22562(1) & -$2.0 \times 10^{-8}$ & \\ \hline
\end{tabular}

\begin{table}
\caption[]{Parameters for a linear fit to the correlation energy
$\bar{\varepsilon}_{xc}$ with $\lambda = 1.0$.}
\label{lambda10_energy}
\end{table}

\begin{tabular}{|c|cll|} \hline
\makebox[1.5cm]{$\nu$} & \makebox[0.5cm]{} & \makebox[2.5cm]{$a_0$} &
\makebox[2.5cm]{$a_1$} \\ \hline
2    & & -4.81968(1) & -$6.26 \times 10^{-7}$  \\
1    & & -4.82005(1) & -$8.45 \times 10^{-8}$  \\
1/3  & & -4.82034(1) & -$2.73 \times 10^{-9}$  \\
1/5  & & -4.82040(1) & -$6.5 \times 10^{-10}$  \\
1/7  & & -4.82042(1) & -$2.3 \times 10^{-10}$  \\ \hline
\end{tabular}

\newpage

\begin{table}
\caption[]{Parameters for a (1,1) rational function fit to the gap
energy $\Delta_{\rm eh}$ for the Coulomb interaction ($\lambda =
0.0$).}
\label{coulomb_gap}
\end{table}

\begin{tabular}{|c|clll|} \hline
\makebox[1.5cm]{$\nu$} & \makebox[0.5cm]{} & \makebox[2.5cm]{$a_0$} &
\makebox[2.5cm]{$a_1$} & \makebox[2.5cm]{$b_1$} \\ \hline
2    & & 0.43(4) & 0.047 & 0.321 \\
1    & & 0.80(8)  & 0.0239 & 0.176 \\
1/3  & & 0.27(3) & 0.00317 & 0.0674 \\
1/5  & & 0.076(8) & -$6.15 \times 10^{-6}$ & 0.0107 \\
1/7  & & 0.040(5) & $3.62 \times 10^{-4}$ & 0.0187 \\ \hline
\end{tabular}

\begin{table}
\caption[]{Parameters for a (1,1) rational function fit to the gap
energy $\Delta_{\rm eh}$ with $\lambda = 0.2$.}
\label{lambda02_gap}
\end{table}

\begin{tabular}{|c|clll|} \hline
\makebox[1.5cm]{$\nu$} & \makebox[0.5cm]{} & \makebox[2.5cm]{$a_0$} &
\makebox[2.5cm]{$a_1$} & \makebox[2.5cm]{$b_1$} \\ \hline
2    & & 0.32(5) & 0.0213 & 0.164 \\
1    & & 0.56(8)  & 0.00735 & 0.0711 \\
1/3  & & 0.13(2) & 0.000847 & 0.0239 \\
1/5  & & 0.046(7) & -0.000177 & 0.0 \\
1/7  & & 0.025(4) & 0.000251 & 0.0155 \\ \hline
\end{tabular}

\newpage

\begin{table}
\caption[]{Parameters for a (1,1) rational function fit to the gap
energy $\Delta_{\rm eh}$ with $\lambda = 0.5$.}
\label{lambda05_gap}
\end{table}

\begin{tabular}{|c|clll|} \hline
\makebox[1.5cm]{$\nu$} & \makebox[0.5cm]{} & \makebox[2.5cm]{$a_0$} &
\makebox[2.5cm]{$a_1$} & \makebox[2.5cm]{$b_1$} \\ \hline
2    & & 0.20(4) & 0.00568 & 0.0472 \\
1    & & 0.30(6)  & 0.00233 & 0.0185 \\
1/3  & & 0.049(9) & 0.000545 & 0.0152 \\
1/5  & & 0.018(4) & -$1.12 \times 10^{-5}$ & 0.0 \\
1/7  & & 0.009(2) & -$6043 \times 10^{-6}$ & 0.0 \\ \hline
\end{tabular}

\begin{table}
\caption[]{Parameters for a (1,1) rational function fit to the gap
energy $\Delta_{\rm eh}$ with $\lambda = 1.0$.}
\label{lambda10_gap}
\end{table}

\begin{tabular}{|c|clll|} \hline
\makebox[1.5cm]{$\nu$} & \makebox[0.5cm]{} & \makebox[2.5cm]{$a_0$} &
\makebox[2.5cm]{$a_1$} & \makebox[2.5cm]{$b_1$} \\ \hline
2    & & 0.14(3) & 0.00248 & 0.0196 \\
1    & & 0.17(4)  & 0.00118 & 0.00819 \\
1/3  & & 0.022(5) & -$3.7 \times 10^{-6}$ & 0.0 \\
1/5  & & 0.008(2) & -$4.4 \times 10^{-7}$ & 0.0 \\
1/7  & & 0.004(1) & -$1.9 \times 10^{-7}$ & 0.0 \\ \hline
\end{tabular}

\end{document}